\documentclass[runningheads]{llncs}
\usepackage{graphicx}

\usepackage{adjustbox}
\usepackage{amsfonts}
\usepackage{amsmath}
\usepackage{booktabs}
\usepackage{hyperref}
\usepackage{lipsum}
\usepackage{mathtools}
\usepackage{multirow}
\usepackage{nicefrac}
\usepackage{soul}
\usepackage{url}

\begin{document}
\title{PauseSpeech: Natural Speech Synthesis via Pre-trained Language Model and Pause-based Prosody Modeling\thanks{This work was supported by Institute of Information \& communications Technology Planning \& Evaluation (IITP) grant funded by the Korea government (MSIT) (No. 2019-0-00079, Artificial Intelligence Graduate School Program (Korea University) and No. 2021-0-02068, Artificial Intelligence Innovation Hub).}}
\titlerunning{Speech Synthesis via Pre-trained Language Model and Pause-based Prosody}
%
\author{Ji-Sang~Hwang \and
Sang-Hoon~Lee \and
Seong-Whan~Lee}
\authorrunning{J.-S. Hwang et al.}
%
\institute{Department of Artificial Intelligence, Korea University, Seoul, Korea\\
\email{\{js\_hwang, sh\_lee, sw.lee\}@korea.ac.kr}\\ }
\maketitle              
\begin{abstract}
Although text-to-speech (TTS) systems have significantly improved, most TTS systems still have limitations in synthesizing speech with appropriate phrasing. For natural speech synthesis, it is important to synthesize the speech with a phrasing structure that groups words into phrases based on semantic information. In this paper, we propose PuaseSpeech, a speech synthesis system with a pre-trained language model and pause-based prosody modeling. First, we introduce a phrasing structure encoder that utilizes a context representation from the pre-trained language model. In the phrasing structure encoder, we extract a speaker-dependent syntactic representation from the context representation and then predict a pause sequence that separates the input text into phrases. Furthermore, we introduce a pause-based word encoder to model word-level prosody based on pause sequence. Experimental results show PauseSpeech outperforms previous models in terms of naturalness. Furthermore, in terms of objective evaluations, we can observe that our proposed methods help the model decrease the distance between ground-truth and synthesized speech. Audio samples are available at \url{https://jisang93.github.io/pausespeech-demo/}.

\end{abstract}
%
%
\section{Introduction}
Text-to-speech (TTS) systems aim to generate high-quality and natural speech from a text. Recently, advancements in generative models \cite{goodfellow2020generative,rezende2015variational} have led to rapid progress in TTS systems to model both linguistic features and variations (e.g., speaker information, prosody, and background noise). Although TTS systems have significantly improved, most TTS systems face limitations in synthesizing speech with proper phrasing structure that groups word into phrases and separates the input text with intentional pauses \cite{klimkov2017phrase}. These limitations usually lead to a disfluent speech when the TTS systems generate an utterance comprising multiple sentences or a long sentence. The disfluent speech conveys too much information at once due to the wrong phrasing structure, resulting in difficult understanding and perceiving it as unnatural by human listeners. Therefore, it is important to synthesize speech with appropriate phrasing based on semantic and syntactic information to enhance comprehensibility \cite{braunschweiler2013automatic} and recall \cite{elmers2021take,klimkov2017phrase}.

\begin{figure}[t]
    \centering
    \includegraphics[width=1.0\textwidth]{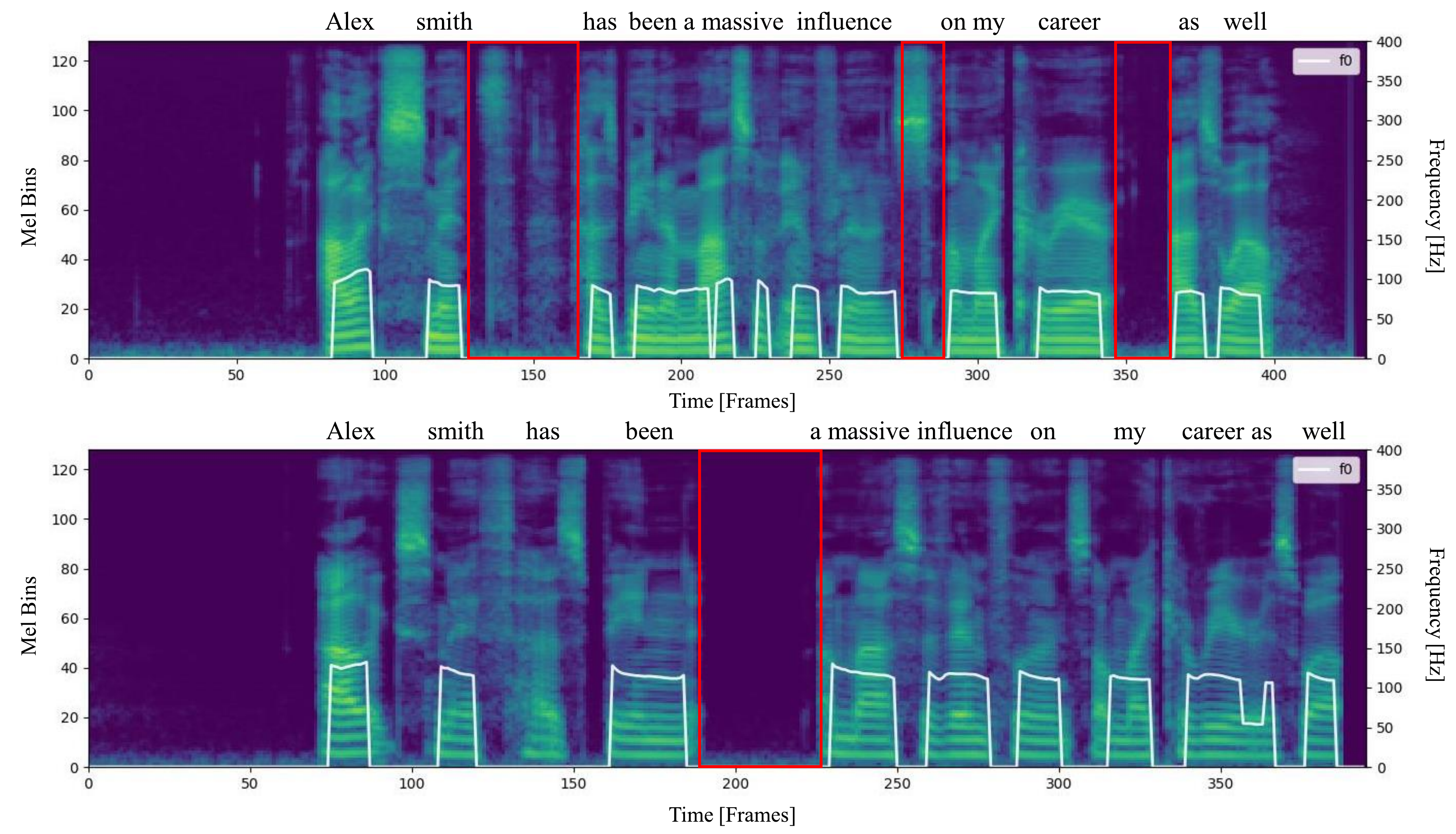}
    \vspace{-0.2cm}
    \caption{Comparison between the Mel-spectrogram of two different speakers and the same text. The corresponding text is ``Alex Smith has been a massive influence on my career as well''. Red boxes represent respiratory pause positions. White lines denote the F0 contour of each utterance.}
\vspace{-0.2cm}
\label{fig:spk_comparison}
\end{figure}

Previous studies \cite{hayashi2019pre,hida2022polyphone,makarov2022simple,xu2021improving} have considered using context information from a pre-trained language model (PLM) \cite{devlin2018bert,lewis2019bart} to enhance the naturalness of generated speech. They have leveraged PLM to improve prosody (e.g., pitch, accent, and prominence) that is related to context information. Additionally, \cite{abbas2022expressive,futamata2021phrase,yang2023duration} have predicted pauses (also known as phrase breaks) based on extracted semantic meanings to enhance the naturalness and comprehensibility of synthesized speech. However, these systems have some limitations: 1) They do not consider that pauses vary according to the speaker. As illustrated in Figure \ref{fig:spk_comparison}, the position and duration of each pause vary from person to person even if it is the same text information. 2) They do not reflect the prosody of surrounding pauses. As patterns of intonation surrounding pauses are slightly different, the systems have to consider the variations surrounding pauses.

To address the aforementioned problems, we propose PauseSpeech, a speech synthesis system utilizing the PLM and pause-based prosody modeling. First, we introduce a phrasing structure encoder using a context representation from the PLM. The phrasing structure encoder encodes the context representation into a speaker-dependent syntactic representation. In the phrasing structure encoder, we predict speaker-dependent pauses using both the syntactic representation and speaker information. Moreover, we propose pause-based prosody modeling to consider the prosody of surrounding pauses in a pause-based word encoder. The pause-based word encoder takes the syntactic representation, a segment-level representation, and word position embedding to extract the pause-based prosody. Experimental results show that PauseSpeech outperforms previous models in terms of naturalness. Furthermore, we can observe in the objective evaluations that our proposed methods help the model decrease the distance between the ground-truth and synthesized speech audio.

\section{Related Works}
\subsection{Text-to-Speech}
Recently, neural TTS systems have significantly improved, resulting in high performance. For speech audio generation, TTS systems predict a pre-defined acoustic feature (e.g., Mel-spectrogram) using an acoustic model \cite{ren2019fastspeech,wang2017tacotron}, which is converted into a waveform by a vocoder \cite{kim2021fre,kong2020hifi}, or directly generate a waveform \cite{donahue2020end,kim2021conditional,lee2022hierspeech}. To predict the pre-defined acoustic features, there are two types of acoustic models: autoregressive (AR) and non-autoregressive (NAR)-TTS systems. AR-TTS systems \cite{shen2018natural,wang2017tacotron} generate each frame of the Mel-spectrogram conditioned on previous frames to model long-term dependency. However, they suffer from slow inference speed and robustness errors, such as word skipping and repetition. NAR-TTS systems \cite{kim2020glow,ren2019fastspeech,lee2022pvae} have been proposed to handle these problems. They map the text sequence into the Mel-spectrogram using alignment between the text and Mel-frames sequence. These systems can generate speech faster and more robustly than AR-TTS systems.

For diverse and expressive speech, TTS systems have adopted auxiliary variation modeling. They use explicit features (e.g., pitch, energy) \cite{lancucki2021fastpitch,ren2020fastspeech,seshadri2021emphasis,lee2021multi} or implicit features \cite{ren2022prosospeech,ren2021portaspeech} to model prosody. Although these prosody modeling can improve diversity and expressiveness, they still have limitations in synthesizing speech with appropriate phrasing structure considering both semantic and syntactic information. In this paper, we use the contextual information from PLM to learn the proper phrasing structure from the input text sequence.

\subsection{Phrase Break Prediction}
Previous TTS systems \cite{abbas2022expressive,futamata2021phrase,klimkov2017phrase,szekely2020breathing,yang2023duration} have considered predicting phrase breaks that can be defined as pauses inserted between phrases to learn proper phrasing structure. Generally, there are two types of pauses: a punctuation-based pause and respiratory pause. The punctuation-based pause is usually generated at a punctuation mark in the text sequence. However, the respiratory pause does not have instructions such as the punctuation mark. Therefore, several TTS systems \cite{futamata2021phrase,klimkov2017phrase} have predicted the respiratory pause and inserted it into space between words. However, generated pauses are still inappropriate due to retrieving the average style in the training dataset.

Several pauses prediction-based TTS systems \cite{abbas2022expressive,futamata2021phrase,yang2023duration} have used contextual representation from PLM. Since the contextual representation contains semantic and syntactic information, the extracted contextual representation helps the systems predict the proper positions of the respiratory pause. Inspired by the previous studies, we adopt the PLM to predict the respiratory pause and utilize it to predict speaker-dependent pauses. Furthermore, we classify the respiratory pauses into four categories following \cite{elmers2021take,yang2023duration}.

\begin{figure}[t]
    \centering
    \includegraphics[width=1.0\textwidth]{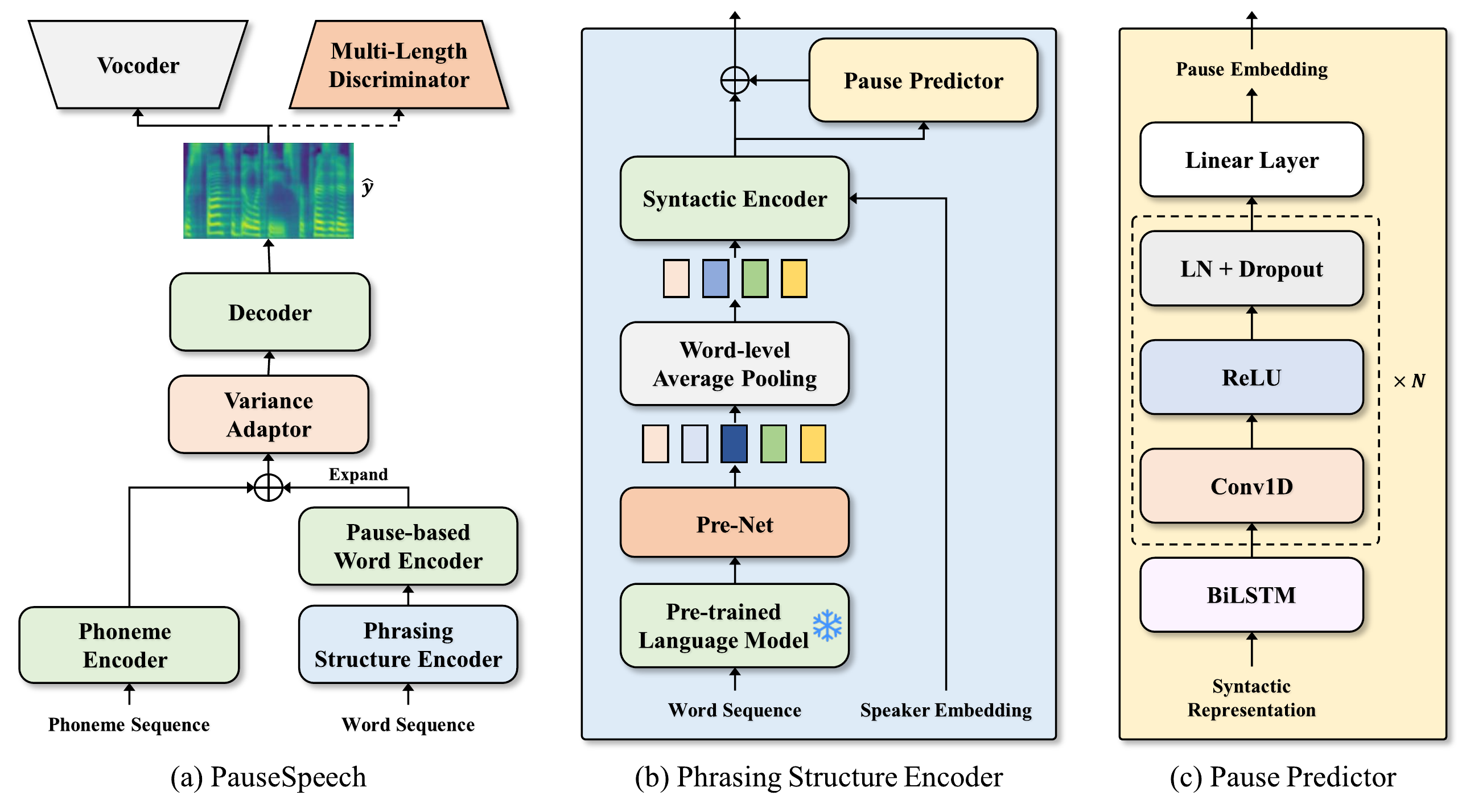}
    \vspace{-0.2cm}
    \caption{(a) Overall architecture of PauseSpeech. The dashed line represents that the operation only utilizes during training. (b) In a phrasing structure encoder, we convert subword-level context representation from PLM into word-level context representation with word-level average pooling. A syntactic encoder takes both the word-level context representation and speaker embedding to extract a speaker-dependent syntactic representation. (c) A pause predictor takes the syntactic representation to predict speaker-dependent pauses.}
\label{fig:arch}
\vspace{-0.2cm}
\end{figure}

\section{PauseSpeech}
In this paper, we propose a TTS system that uses a pre-trained language model (PLM) and pause-based prosody modeling for natural and expressive speech. We propose a phrasing structure encoder using contextual representation from the PLM to extract syntactic representation and predict pauses. Moreover, we introduce pause-based prosody modeling in a pause-based word encoder to consider explicit variations surrounding pauses. Furthermore, we adopt adversarial learning to enhance the quality of the generated Mel-spectrogram. We describe the details of PauseSpeech in the following subsection.

\subsection{Phrasing Structure Encoder}
We introduce the phrasing structure encoder that encodes the context representation and speaker information into speaker-dependent syntactic representation. The phrasing structure encoder comprises a pre-net, syntactic encoder, and pause predictor as illustrated in Figure \ref{fig:arch} (b).

\subsubsection{Syntactic encoder}
We use BERT \cite{devlin2018bert} as the PLM to extract context representation. It is well known that the context representation from BERT contains syntactic information and knowledge of semantic roles \cite{liu2019roberta,rogers2021primer}. In particular, previous studies \cite{hewitt2019structural,liu2019linguistic} demonstrated that the context representation from the middle layer of BERT contained more prominent syntactic and semantic information than other layers. Therefore, we utilize the syntactic and semantic information by extracting the self-supervised context representation from the input text sequence.

The syntactic encoder is designed to extract the speaker-dependent syntactic representation. We supposed that the human speaker's text cognition varies from person to person, resulting in varying the position and duration of the respiratory pause in human speech. Therefore, the syntactic encoder takes both the context representation and speaker embedding to obtain the syntactic representation that contains syntactic information based on the target speaker's cognition. As BERT extracts the input text sequence into subword-level contextual representation, we process the context representation with word-level average pooling \cite{ren2021portaspeech,ye2022syntaspeech} to convert the representation into a word-level sequence.

\subsubsection{Pause predictor}
\label{subsub:pause_pred}
We define categories of the respiratory pause according to the pause duration to classify the pause as an intentional/unintentional pause. We use the Montreal forced aligner \cite{mcauliffe2017montreal} to obtain the pause duration. Following previous studies \cite{elmers2021take,yang2023duration}, we categorize the pause into four-class: no pause ($0-100$ ms), short pause ($100-300$ ms), medium pause ($300-700$ ms), and long pause ($> 700$ ms). Moreover, we label each pause class as follows: ``0'' denotes the no pause, and ``1'', ``2'', and ``3'' represents the short, medium, and long pause, respectively. In this paper, we also define the short pause as the unintentional pause and the medium and long pauses as the intentional pause.

The pause predictor takes the syntactic representation to predict the speaker-dependent pause sequence. The pause predictor comprises two Bi-LSTM layers and 1D-convolutional networks with ReLU activation, layer normalization, and dropout as shown in \ref{fig:arch} (c). The final linear layer projects hidden representation into a word-level pause sequence. We encode the pause sequence into trainable pause embedding ad add it to the output of the syntactic encoder. Furthermore, we optimize the pause predictor with a cross-entropy loss between a probability distribution with a softmax function and a target pause label sequence. 

\begin{figure}[t]
    \centering
    \includegraphics[width=1.0\textwidth]{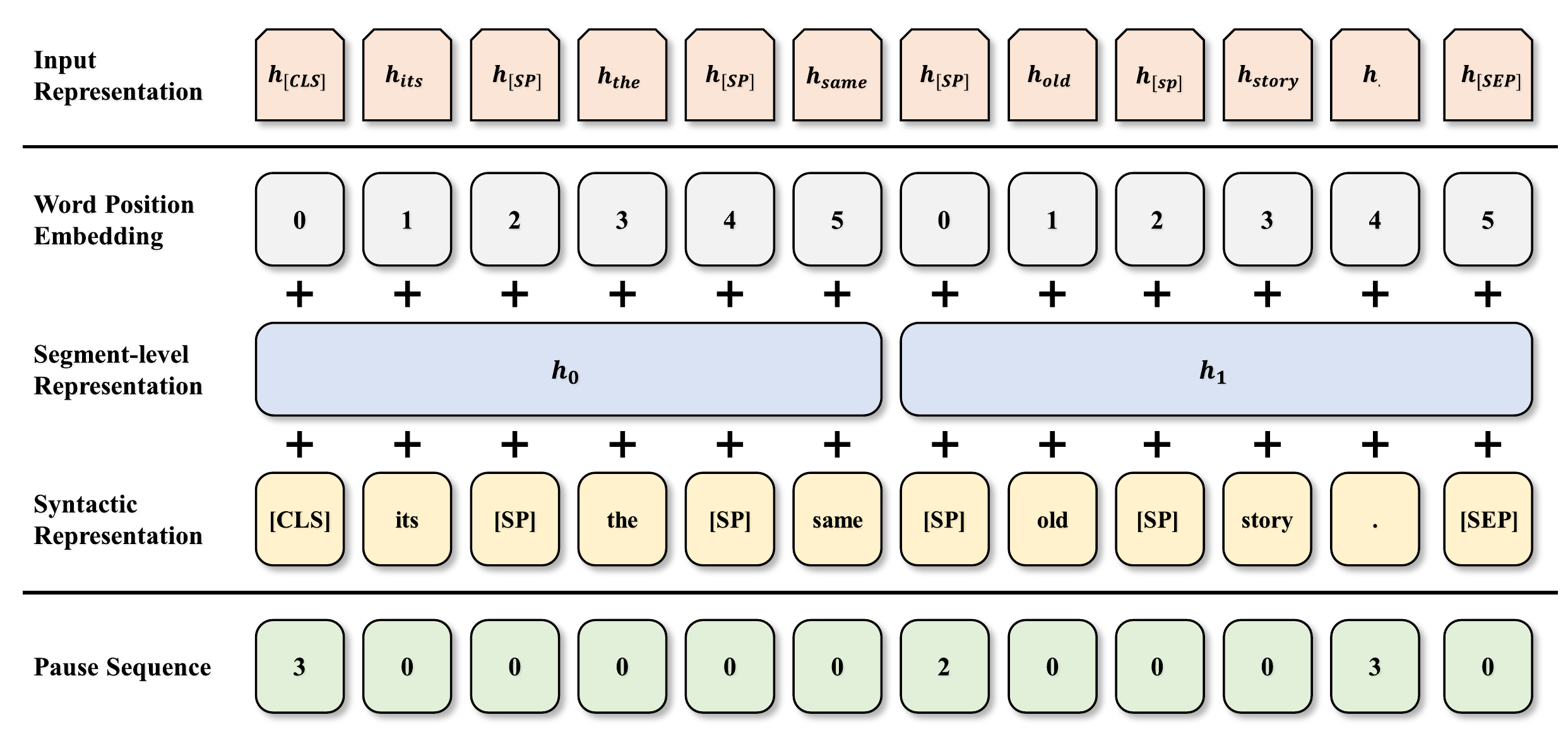}
    \vspace{-0.2cm}
    \caption{Input representation for pause-based word encoder. The input representations are the sum of the word-level syntactic representation, segment-level representation, and segment-word position embedding.}
\label{fig:pause-modeling}
\vspace{-0.2cm}
\end{figure}

\subsection{Pause-Based Word Encoder}
We introduce a pause-based word encoder to model a pause-based prosody that considers the prosody of word sequence surrounding pauses. For pause-based prosody modeling, inputs to the pause-based word encoder comprise three components as illustrated in Figure \ref{fig:pause-modeling}: an output of the phrasing structure encoder, a segment-level representation, and word position embedding. The first component is the summation of the output of the syntactic encoder and pause embedding in the phrasing structure encoder. For the second component, we process the output of the phrasing structure encoder with segment-level average pooling. For segmentation, we divide the word sequence into segments with the intentional pause (medium and long pause), which is defined in subsection \ref{subsub:pause_pred}. Additionally, we design that the punctuation-based pause would be assigned to a previous segment. For the third component, we provide position information within each segment. We implement the same sinusoidal function for the word position embedding, which is used in BERT.

\subsection{Adeversarial Learning}
The generated Mel-spectrogram from TTS systems generally suffered blurry and over-smoothing problems due to using simple optimization functions such as mean absolute error (MAE), and mean square error (MSE) \cite{ren2022revisiting,ye2022syntaspeech}. Following \cite{wu2020adversarially,ye2022syntaspeech}, we adopt a multi-length discriminator to improve the quality of the generated Mel-spectrogram. The multi-length discriminator distinguishes between the generated and ground-truth Mel-spectrogram, which are randomly sliced by windows of multi-lengths. Moreover, for robust training, we solely train the generator until 50K steps, and then jointly train with the multi-length discriminator after 50K steps.

\section{Experiment and Results}
\subsection{Experimental Setup}
\subsubsection{Datasets}
We trained PauseSpeech on VCTK\footnote{\url{https://datashare.ed.ac.uk/handle/10283/3443}} \cite{veaux2016superseded} to synthesize speech. The VCTK dataset contains approximately 46 hours of audio for 108 English speakers. We divided the VCTK dataset into three subsets: 40,857 samples for training, 1,500 samples for validation, and 1,500 samples for testing.

In addition, we downsampled the audio at 24,000 Hz for training. We transform the raw waveform into the Mel-spectrogram with 128 bins. For the input of the phoneme encoder, we converted the text sequence into the phoneme sequence using the open-source grapheme-to-phoneme tool\footnote{\url{https://github.com/Kyubyong/g2p}}.

\subsubsection{Model configuration}
PauseSpeech consists of a phoneme encoder, phrasing structure encoder, pause-based word encoder, variance adaptor, decoder, and multi-length discriminator. The phoneme and pause-based word encoders and the decoder comprise four feed-forward Transformer (FFT) blocks \cite{ren2019fastspeech} with relative-position encoding \cite{shaw2018self} following Glow-TTS \cite{kim2020glow}. For the variance adaptor, we adopt the same architecture in FastSpeech 2 \cite{ren2020fastspeech}. For adversarial learning, we adopt the multi-length discriminator of SyntaSpeech \cite{ye2022syntaspeech}, which comprises stacked convolutional layers with batch normalization.

In the phrasing structure encoder, we use BERT-base model\footnote{\url{https://huggingface.co/bert-base-uncased}} for the PLM. We utilized the 9th layer of BERT to extract the self-supervised contextual representation. The phrasing structure encoder consists of a pre-net, syntactic encoder, and pause predictor. The pre-net comprises two BiLSTM layers and multiple stacked convolutional layers with ReLU activation, layer normalization, and dropout. The syntactic encoder has the same architecture as the other encoder, which consists of four FFT blocks with relative-position encoding.

\subsubsection{Training}
We trained PauseSpeech using the Adam optimizer \cite{loshchilov2017decoupled} with a learning rate of $2 \times 10^{-4}$, $\beta_{1}=0.8$, $\beta_{2}=0.99$, and weight deacy of $\lambda=0.01$. PauseSpeech has been trained on two NVIDIA RTX A6000 GPUs with 32 sentences per GPU. It takes 400K steps for training until convergence. In addition, we use pre-trained HiFi-GAN \cite{kong2020hifi} as the vocoder to convert the synthesized Mel-spectrogram into raw waveform.

\subsection{Evaluation Metrics}
\subsubsection{Subjective metrics}
We conducted the mean opinion score (MOS) evaluation on the test dataset to evaluate the naturalness of the audio via Amazon Mechanical Turk. The MOS test was rated by at minimum of 30 listeners on a scale of 1-5. The MOS evaluation is reported with 95\% confidence intervals.

\subsubsection{Objective metrics}
We calculated various types of distance between the ground-truth and synthesized audio. We used five objective metrics to evaluate the quality of synthesized speech: 1) Phoneme error rate (PER); 2) Word error rate (WER); 3) Mel-cepstral distortion (MCD); 4) F0 root mean square error ($\text{RMSE}_\text{F0}$); 5) Average absolute difference of the utterance duration (DDUR) \cite{zhang2019sequence}. For PER and WER evaluations, we used the open-source automatic speech recognition (ASR) model\footnote{\url{https://huggingface.co/facebook/wav2vec2-large-960h-lv60-self}}, which is trained over wav2vec 2.0 \cite{baevski2020wav2vec}. We calculated PER and WER between the ASR prediction and target text. For MCD and $\text{RMSE}_{\text{F0}}$ evaluations, we applied dynamic time warping between the ground-truth and synthesized audio.

\begin{table}[t]
\centering
\caption{Performance comparison with different methods. Recon. represents reconstruction.}
\resizebox{1.0\textwidth}{!}{
  \begin{tabular}{l|c|c c|c c c}
  \toprule
     Method & MOS ($\uparrow$) &$\ $ PER ($\downarrow$) & WER ($\downarrow$) $\ $&$\ $ MCD ($\downarrow$) & $\text{RMSE}_\text{F0} $ ($\downarrow$) & DDUR ($\downarrow$) $\ $\\
  \midrule
     GT & $3.91 \pm 0.03$ & $1.41$ & $3.73$ & $-$ & $-$ & $-$\\
     HiFi-GAN (recon.)$\ $ & $3.90 \pm 0.03$ & $2.01$ & $4.78$ & $0.94$ & $21.96$ & $-$\\
  \midrule
     FastSpeech 2 & $3.81 \pm 0.03$ & $2.43$ & $5.17$ & $3.60$ & $32.18$ & $0.16$\\
     PortaSpeech & $3.83 \pm 0.03$ & $2.06$ & $4.43$  & $3.55$ & $33.72$ & $\mathbf{0.13}$\\
  \midrule
     PauseSpeech & $\ \mathbf{3.88 \pm 0.03}\ $ & $\mathbf{1.32}$ & $\mathbf{3.44}$ & $\mathbf{3.42}$ & $\mathbf{27.66}$ & $\mathbf{0.13}$\\
  \bottomrule
  \end{tabular}
}\vspace{-0.2cm}
\label{table:MOS_comparison}
\end{table}

\subsection{Performance}
We compared the audio generated by PauseSpeech with the outputs of the following systems: 1) GT, the ground-truth audio; 2) HiFi-GAN \cite{kong2020hifi}, where we reconstructed the audio from the ground-truth Mel-spectrogram using the pre-trained vocoder; 3) FastSpeech 2 \cite{ren2020fastspeech}, 4) PortaSpeech \cite{ren2021portaspeech}. We converted the synthesized Mel-spectrogram from the acoustic models into a raw waveform using the pre-trained HiFi-GAN.

The results are shown in Table \ref{table:MOS_comparison}. We observed that PauseSpeech outperformed the previous systems in terms of naturalness. Moreover, our proposed model significantly reduced the PER and WER. This indicates that PauseSpeech generates speech with accurate pronunciation. Furthermore, our model achieved better performance in terms of MCD and $\text{RMSE}_{F0}$. These observations suggest that PauseSpeech can reduce the distance between the ground-truth and synthesized audio.

\begin{figure}[ht]
    \centering
    \includegraphics[width=1.0\textwidth]{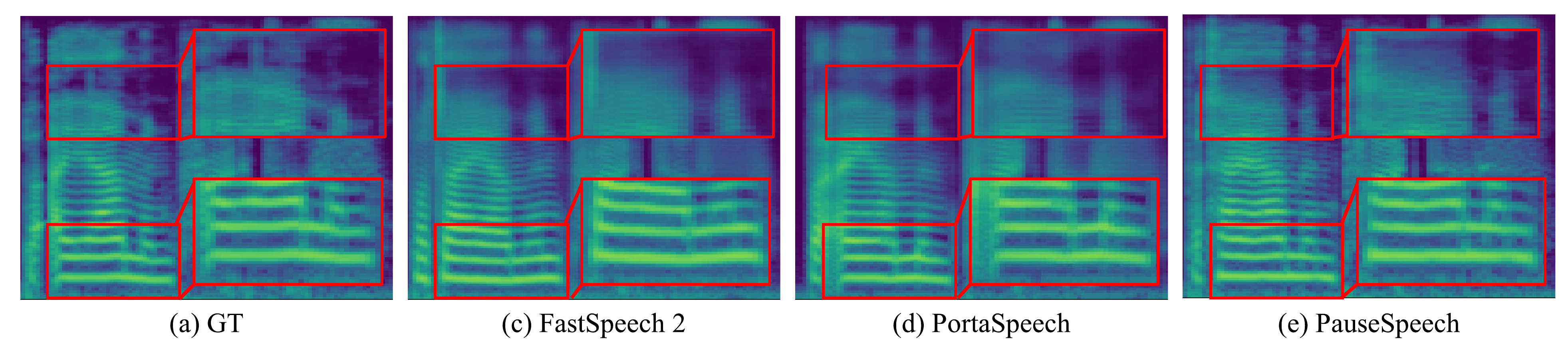}
    \vspace{-0.2cm}
    \caption{Visualization of the Mel-spectrogram with varying systems: (a) GT, (b) FastSpeech 2, (c) PortaSpeech, and (d) PauseSpeech. The corresponding text is ``But I am in practice''.}
\label{fig:mel_compare}
\vspace{-0.2cm}
\end{figure}

We also visualized the Mel-spectrograms of synthesized audio from the varying systems to compare the systems in Figure \ref{fig:mel_compare}. In the low-frequency bands of the Mel-spectrogram, we can observe that PauseSpeech generates a similar pitch contour, resulting in expressive prosody. Moreover, PauseSpeech generates more details in the high-frequency bands of the Mel-spectrogram, resulting in natural sounds. These results demonstrate that PauseSpeech can generate high-quality and natural speech audio with accurate pronunciation.

\begin{table}[ht]
\centering
\caption{Experimental results on the context representations from each different layer of BERT.}
\resizebox{1.0\textwidth}{!}{
  \begin{tabular}{l|c|c c|c c c}
  \toprule
     Layer & MOS ($\uparrow$) &$\ $ PER ($\downarrow$) & WER ($\downarrow$) $\ $&$\ $ MCD ($\downarrow$) & $\text{RMSE}_\text{F0} $ ($\downarrow$) & DDUR ($\downarrow$) $\ $\\
  \midrule
     Lower layer (1st) & $3.91 \pm 0.03$ & $1.43$ & $3.56$ & $\mathbf{3.42}$ & $26.66$ & $0.14$\\
     Middle layer (9th) & $\ \mathbf{3.98 \pm 0.03}\ $ & $\mathbf{1.32}$ & $\mathbf{3.44}$ & $\mathbf{3.42}$ & $27.66$ & $\mathbf{0.13}$\\
     Higher layer (12th)$\ $ & $3.93 \pm 0.03$ & $1.35$ & $3.59$ & $3.44$ & $\mathbf{26.59}$ & $\mathbf{0.13}$\\
  \bottomrule
  \end{tabular}
}\vspace{-0.2cm}
\label{table:self-supervised}
\end{table}

\subsection{Analysis of Self-supervised Representations}
Previous studies \cite{hewitt2019structural,liu2019linguistic} have shown that the middle layer of BERT contains the most prominent syntactic information. In addition, \cite{goldberg2019assessing} demonstrated that the 8th and 9th layers of pre-trained BERT showed the best subject-verb agreement. Therefore, we divided the layers of BERT into three parts to verify the effectiveness of each part in TTS tasks and selected one layer of each part: the 1st layer of the lower layer, the 9th layer of the middle layer, and the 12th layer of the higher layer. Then, we extracted the self-supervised representation from each layer as an input of the phrasing structure encoder to compare the performance.

As shown in Table \ref{table:self-supervised}, the representation of the 9th layer of BERT has better performance than the others in terms of naturalness. Moreover, the representation of the 9th layer of BERT showed the lowest error distance in the most of objective evaluations. In particular, the representation of the 9th layer of BERT significantly decreased PER and WER. These results demonstrate that the middle layer of BERT contains rich information, resulting in improvement in the performance of the TTS system. Additionally, the higher subject-verb agreement may help the TTS model synthesize speech with accurate pronunciation. Hence, we used the representation from the 9th layer of BERT for the phrasing structure encoder.

\begin{table}[h]
\centering
\caption{Ablation study of PauseSpeech. PW and PS encoder denotes the pause-based word encoder and the phrasing structure encoder, respectively. Adv. learning represents adversarial learning.}
\resizebox{0.80\textwidth}{!}{
  \begin{tabular}{l|c|c c c}
  \toprule
     Layer & MOS ($\uparrow$) &$\ $ MCD ($\downarrow$) & $\text{RMSE}_\text{F0} $ ($\downarrow$) & DDUR ($\downarrow$) $\ $\\
  \midrule
     PauseSpeech & $\ \mathbf{3.98 \pm 0.03}\ $ & $\mathbf{3.42}$ & $27.66$ & $\mathbf{0.13}$\\
  \midrule
     w/o PW encoder & $3.94 \pm 0.03$ & $3.46$ & $\mathbf{27.64}$ & $\mathbf{0.13}$\\
     $\quad$ w/o PS encoder$\ $ & $3.93 \pm 0.03$ & $3.50$ & $28.94$ & $0.14$\\
  \midrule
     w/o adv. learning & $3.92 \pm 0.03$ & $3.48$ & $33.48$ & $0.14$\\
  \bottomrule
  \end{tabular}
}\vspace{-0.2cm}
\label{table:abl}
\end{table}

\subsection{Ablation Study}
We conducted an ablation study to demonstrate the effectiveness of each module in PauseSpeech. We compared PauseSpeech with that without the pause-based word encoder and that without the phrasing structure encoder. The results are presented in Table \ref{table:abl}. We observed that removing encoders degraded the naturalness of synthesized audio. Moreover, PauseSpeech without the pause-based word and phrasing structure encoders significantly degraded the objective evaluations. These results indicate that our proposed modules are necessary to synthesize natural speech. Furthermore, we trained PauseSpeech without adversarial learning. We observed that removing the adversarial learning significantly degraded the performance. These results indicate that adversarial learning enhances the performance of the system, resulting in synthesizing high-quality and natural speech.

\section{Conclusions}
In this study, we presented a TTS model, PauseSpeech, which can learn and synthesize speech with a proper phrasing structure using a pre-trained language model and pause-based prosody modeling. PauseSpeech uses the contextual representation from the pre-trained BERT and then models pause-based prosody based on predicted pauses. We also adopted a multi-length discriminator for adversarial learning. Our experimental results show that PauseSpeech outperforms previous TTS models in terms of naturalness and significantly enhances the pronunciation of synthesized speech. We also conducted ablation studies to verify the effectiveness of each component in PauseSpeech. In future works, we will verify the effectiveness of PauseSpeech in multilingual scenarios and attempt to control more diverse variations of speech. 

%
%
\bibliographystyle{splncs04}
\bibliography{reference.bib}
\end{document}